%% file: PolarCodes.tex
\documentclass[a4paper,journal]{IEEEtran}
\usepackage{graphicx,ifthen}
\usepackage[cmex10]{amsmath}
\usepackage{psfrag,amssymb,amsthm}
\usepackage{mathtools}

\usepackage[noadjust]{cite}
\usepackage{color,pst-plot,stfloats,pst-sigsys}
\usepackage{pstricks-add}
\usepackage{subfigure}

\newtheorem{lem}{Lemma}
\newtheorem{cor}{Corollary}
\newtheorem{prop}{Proposition}

\theoremstyle{definition}
\newtheorem{definition}{Definition}

\newtheorem{example}{Example}

\title{The Fractality of Polar and Reed-Muller Codes}

\author{Bernhard C. Geiger,~\IEEEmembership{Member, IEEE}
\thanks{Parts of this work have been presented at the 2015 NEWCOM\# Emerging Topics in Modulation and Coding Workshop and will be presented at the 2016 International Z\"urich Seminar on Communications.}
\thanks{The work of Bernhard C. Geiger has been funded by the Erwin Schr\"odinger Fellowship J 3765 of the Austrian Science Fund and by the German Ministry of Education and Research in the framework of an Alexander von Humboldt Professorship.}
\thanks{Bernhard C. Geiger (geiger@ieee.org) is with the Institute for Communications Engineering, Technical University of Munich.}
} 

\input{abbrevations_processing.tex}

\newcommand{\Gset}{\mathcal{G}}
\newcommand{\Bset}{\mathcal{B}}
\newcommand{\Fset}{\mathcal{F}}
\newcommand{\Hset}{\mathcal{H}}
\renewcommand{\Prob}{\mathbb{P}}

\begin{document}

\maketitle

\begin{abstract}
The generator matrices of polar codes and Reed-Muller codes are obtained by selecting rows from the Kronecker product of a lower-triangular binary square matrix. For polar codes, the selection is based on the Bhattacharyya parameter of the row, which is closely related to the error probability of the corresponding input bit under sequential decoding. For Reed-Muller codes, the selection is based on the Hamming weight of the row. This work investigates the properties of the index sets pointing to those rows in the infinite blocklength limit. In particular, the Lebesgue measure, the Hausdorff dimension, and the self-similarity of these sets will be discussed. It is shown that these index sets have several properties that are common to fractals.
\end{abstract}

\begin{IEEEkeywords}
Polar codes, Reed-Muller codes, fractals, self-similarity
\end{IEEEkeywords}

\section{Introduction}\label{sec:intro}
Polar codes and Reed-Muller codes are Kronecker product-based codes. Such a code of block-length $2^n$ is based on the $n$-fold Kronecker product $G(n):=F^{\otimes n}$, where
\begin{equation}
 F:=\left[\begin{array}{cc}
           1 & 0\\ 1 & 1
          \end{array}
\right].\label{eq:Fmatrix}
\end{equation}
Following the terminology of~\cite{Kahraman_Fractal}, a rate-$K/2^n$ Kronecker product-based code is uniquely defined by a set $\Fset$ of $K$ indices: Its generator matrix is the submatrix of $G(n)$ consisting of the rows indexed by $\Fset$. For polar codes~\cite{Arikan_PolarCodes}, in which each row of $G(n)$ can be interpreted as a (partially polarized) channel, $\Fset$ consists of rows corresponding to the $K$ channels with the lowest Bhattacharyya parameters~\cite{Arikan_PolarRM} (the ``good'' channels, see Section~\ref{sec:prelim:polar}). For Reed-Muller codes, $\Fset$ consists of those rows of $G(n)$ with a Hamming weight above a certain threshold (see Section~\ref{sec:prelim:RM}). Despite its importance for code construction, at least for polar codes, very little is known about the structure of $\Fset$. A recent exception is the work by Renes, Sutter, and Hassani, stating conditions under which polarized sets are aligned, i.e., under which the good (bad) channels derived from one binary-input memoryless channel are a subset of the good (bad) channels derived from another~\cite{Renes_Aligned}.

That Kronecker product-based codes, such as polar codes~\cite{Arikan_PolarCodes} or Reed-Muller codes, possess a fractal nature has been observed in~\cite{Kahraman_Fractal}, noting the similarity between $G(n)$ and the Sierpinski triangle. Much earlier, Abbe suspected that the set of ``good''  polarized channels is fractal~\cite{Abbe_PC}. Nevertheless, to the best of the author's knowledge, no definite statement regarding this fractal nature has been made yet. In this paper, we try to fill this gap and present results about the sets $\Fset$ for polar codes (Section~\ref{sec:polar}) and Reed-Muller codes (Section~\ref{sec:reedmuller}). The self-similar structure of these sets is also suggested in~\cite{Bardet_DecreasingMonomial}, which shows that polar and Reed-Muller codes are decreasing monomial codes. While~\cite{Bardet_DecreasingMonomial} focuses on finite blocklengths, we study the properties of $\Fset$ for infinite blocklengths, i.e., for $n\to \infty$.

To simplify analysis, we represent every infinite binary sequence indexed in $\Fset$ by a point in the unit interval $[0,1]$. Let $\Omega:=\{0,1\}^\infty$ be the set of infinite binary sequences, and let $b:=(b_1b_2\cdots)\in\Omega$ be an arbitrary such sequence. We abbreviate $b^n:=(b_1b_2\cdots b_n)$. Let $(\Omega,\mathfrak{B},\Prob)$ be a probability space with $\mathfrak{B}$ the Borel field generated by the cylinder sets $S(b^n):=\{w\in\Omega{:}\ w_1=b_1,\dots,w_n=b_2\}$ and $\Prob$ a probability measure satisfying $\Prob(S(b^n))=1/2^n$. The following function $f{:}\ \Omega\to [0,1]$ converts these sequences to real numbers:
\begin{equation}
 f(b):=\sum_{n=1}^\infty \frac{b_n}{2^n}\label{eq:functionf}
\end{equation}

Letting $\mathbb{D}:=[0,1]\cap \{p/2^n{:}\ p\in\mathbb{Z},n\in\mathbb{N}\}$ denote the set of dyadic rationals in the unit interval, we recognize that $f$ is non-injective:
\begin{example}\label{ex:dyadics}
 $f$ maps both $b=(01111111\cdots)$ and $b=(10000000\cdots)$ to $0.5$. We call the latter binary expansion \emph{terminating}.
\end{example}

However, as the following lemma shows, $f$ is bijective if we exclude the dyadic rationals:
\begin{lem}[{\cite[Exercises 7-10, p.~80]{Taylor_MeasureTheory}}]\label{lem:fprops}
 Let $\mathfrak{B}_{[0,1]}$ be the Borel $\sigma$-algebra on $[0,1]$ and let $\lambda$ be the Lebesgue measure. Then, the function $f$ in~\eqref{eq:functionf} satisfies the following properties:
 \begin{enumerate}
 \item $f$ is measurable w.r.t.\ $\mathfrak{B}_{[0,1]}$
 \item $f$ is bijective on $\Omega\setminus f^{-1}(\mathbb{D})$
 \item for all $I\in\mathfrak{B}_{[0,1]}$, $\Prob(f^{-1}(I)) = \lambda(I)$
\end{enumerate}
\end{lem}

We believe that the results we prove in the following not only improve our understanding of polar and Reed-Muller codes: Since its introduction in 2009, the polarization technique proposed by Ar{\i}kan has found its way into areas different from polar coding. Haghighatshoar and Abbe showed in the context of compression of analog sources that R\'{e}nyi information dimension can be polarized~\cite{Haghighatshoar_RenyiPolarization}, and Abbe and Wigderson used polarization for the construction of high-girth matrices~\cite{Abbe_HighGirth}. Recently, Nasser proved that a binary operation is polarizing if and only if it is uniformity preserving and its inverse is strongly ergodic~\cite{Nasser_Ergodic1,Nasser_Ergodic2}. We believe that our results might carry over to these areas as well; Section~\ref{sec:discussion} points to possible extensions.

\section{Preliminaries for Polar Codes}\label{sec:prelim:polar}
We adopt the notation of~\cite{Arikan_PolarCodes}. Let $W{:}\ \{0,1\}\to\dom{Y}$ be a binary-input memoryless channel with output alphabet $\dom{Y}$, capacity $0<I(W)<1$, and with Bhattacharyya parameter
\begin{equation}
 Z(W) := \sum_{y\in\dom{Y}} \sqrt{W(y|0)W(y|1)}.
\end{equation}
That $Z(W)=0\Leftrightarrow I(W)=1$ and $Z(W)=1\Leftrightarrow I(W)=0$ is a direct consequence of~\cite[Prop.~1]{Arikan_PolarCodes}. We say a channel is \emph{symmetric} if there exists a permutation $\pi{:}\ \dom{Y}\to\dom{Y}$ such that $\pi^{-1}=\pi$ and, for every $y\in\dom{Y}$, $W(y|0)=W(\pi(y)|1)$.

The heart of Ar{\i}kan's polarization technique is that two channel uses of $W$ can be \emph{combined and split} into one use of a ``worse'' channel
\begin{subequations}\label{eq:channels}
\begin{equation}
 W_2^0(y_1^2|u_1) := \frac{1}{2} \sum_{u_2} W(y_1|u_1\oplus u_2) W(y_2|u_2)\label{eq:worse}
\end{equation}
and one use of a ``better'' channel
\begin{equation}
 W_2^1(y_1^2,u_1|u_2) := \frac{1}{2} W(y_1|u_1\oplus u_2) W(y_2|u_2)\label{eq:better}
\end{equation}
\end{subequations}
where $u_1,u_2\in\{0,1\}$ and $y_1,y_2\in\dom{Y}$. In essence, the combining operation codes two input bits by $F$ in~\eqref{eq:Fmatrix} and transmits the coded bits over $W$ via two channel uses, creating a vector channel. The splitting operation splits this vector channel into the two virtual binary-input memoryless channels indicated in~\eqref{eq:channels}. Of these, the better (worse) channel has a strictly larger (smaller) capacity than the original channel $W$, i.e., $I(W_2^0)<I(W)<I(W_2^1)$, while the sum capacity equals twice the capacity of the original channel, i.e., $I(W_2^0)+I(W_2^1)=2I(W)$~\cite[Prop.~4]{Arikan_PolarCodes}.

The effect of combining and splitting on the channel capacities $I(W_2^0)$ and $I(W_2^1)$ admits no closed-form expression; the effect on the Bhattacharyya parameter at least admits bounds:
\begin{lem}[{\cite[Prop.~5~\&~7]{Arikan_PolarCodes}}]\label{lem:Bhattacharyya}
\begin{subequations}
  \begin{align}
  Z(W_2^1) &= g_1(Z(W)) := Z^2(W)< Z(W)\\
  Z(W)<Z(W_2^0) &\le g_0(Z(W)) := 2Z(W)-Z^2(W) \label{eq:Bhattacharyya:worse}
 \end{align}
\end{subequations}
with equality if $W$ is a binary erasure channel.
\end{lem}

Channels with larger blocklengths $2^n$, $n>1$, can either be obtained by direct $n$-fold combining (using the matrix $G(n)$) and $n$-fold splitting, or by recursive pairwise combining and splitting. For $b^n\in\{0,1\}^n$, we obtain
\begin{equation}
 \left(W_{2^n}^{b^n},W_{2^n}^{b^n}\right) \to \left(W_{2^{n+1}}^{b^n0},W_{2^{n+1}}^{b^n1}\right)
\end{equation}
where $b^n0$ and $b^n1$ denote the sequences of zeros and ones obtained by appending $0$ and $1$ to $b^n$, respectively. Note that $g_1$ and $g_0$ from Lemma~\ref{lem:Bhattacharyya} are non-negative and non-decreasing functions mapping the unit interval onto itself, hence the inequality in~\eqref{eq:Bhattacharyya:worse} is preserved under composition:
\begin{equation}
 Z(W_{2^n}^{b^n}) \le p_{b^n}(Z(W)):=g_{b_n}(g_{b_{n-1}}(\cdots g_{b_1}(Z(W)) \cdots ))
\end{equation}

The channel polarization theorem shows that, with probability one, after infinitely many combinations and splits, only perfect or useless channels remain, i.e., either $I(W_\infty^{b})=1$ or $I(W_\infty^{b})=0$ for $b\in\{0,1\}^\infty$. This is made precise in: 

\begin{prop}[{\cite[Prop.~10]{Arikan_PolarCodes}}]\label{prop:polarization}
 With probability one, the limit RV $I_\infty(b):=I(W_\infty^b)$ takes values in the set $\{0,1\}$: $\Prob(I_\infty=1)=I(W)$ and $\Prob(I_\infty=0)=1-I(W)$.
\end{prop}

This immediately gives rise to
\begin{definition}[The Good and the Bad Channels]\label{def:gset}
 Let $\Gset$ denote the set of good channels, i.e.,
 \begin{subequations}
   \begin{equation}
  x\in\Gset \Leftrightarrow \exists b\in f^{-1}(x){:}\ I(W_\infty^b)=1.
 \end{equation}
 Let $\Bset$ denote the set of bad channels, i.e.,
  \begin{equation}
  x\in\Bset \Leftrightarrow \exists b\in f^{-1}(x){:}\ I(W_\infty^b)=0.
 \end{equation}
 \end{subequations}
\end{definition}

If the polarization procedure is stopped at a finite blocklength $2^n$ for $n$ large enough, it can still be shown that the vast majority of the resulting $2^n$ channels are either almost perfect or almost useless, in the sense that the channel capacities are close to one or to zero (or that the corresponding Bhattacharyya parameters are close to zero or to one). The idea of polar coding is to transmit data only on those channels that are almost perfect: $n$-fold combining and splitting leads to $2^n$ virtual channels, each corresponding to a row of $G(n)$. The channels with high capacity are indicated by $\Fset$, and the generator matrix of the corresponding polar code is the submatrix of $G(n)$ consisting of those indicated rows. If the blocklength grows to infinity ($n\to\infty$), the set $\Fset$ becomes equivalent to the set $\Gset$ in Definition~\ref{def:gset}.

The difficulty of polar coding lies in code construction, i.e., in determining \emph{which} channels/row indices are in the sets $\Fset$ and $\Gset$ for finite and infinite blocklengths. This immediately translates to the question which sequences $b\in\{0,1\}^\infty$ correspond to combinations and splits leading to a perfect channel (or which finite-length sequences $b^n$ lead to channels with capacity sufficiently close to one). Determining the capacity of the virtual channels is an inherently difficult operation, since, whenever $W$ is not a binary erasure channel (BEC), the cardinality of the output alphabet increases exponentially in $2^n$~\cite[Ch.~3.3]{Sasoglu_PolarCodes},~\cite[p.~36]{Korada_PhD}. To circumvent this problem, Tal and Vardy presented an approximate construction method in~\cite{Tal_PolarConstruction}, that relies on reduced output alphabet channels that are either upgraded or degraded w.r.t.\ the channel of interest. As these upgrading/degrading properties -- mentioned earlier in Korada's PhD thesis~\cite[Def.~1.7 \& Lem.~1.8]{Korada_PhD} -- play a fundamental role in this work, we present

\begin{definition}[Channel Up- and Degrading]
 A channel $W^-{:}\ \{0,1\}\to\dom{Z}$ is \emph{degraded} w.r.t.\ the channel $W$ (short:  $W^- \preccurlyeq W$) if there exists a channel $Q{:}\ \dom{Y}\to\dom{Z}$ such that
 \begin{equation}
  W^-(z|u) = \sum_{y\in\dom{Y}} W(y|u)Q(z|y).
 \end{equation}
 A channel $W^+{:}\ \{0,1\}\to\dom{Z}$ is \emph{upgraded} w.r.t.\ the channel $W$ (short:  $W^+ \succcurlyeq W$) if there exists a channel $P{:}\ \dom{Z}\to\dom{Y}$ such that
 \begin{equation}
  W(y|u) = \sum_{z\in\dom{Z}} W^+(z|u)P(y|z).
 \end{equation}
 Moreover, $W^+ \succcurlyeq W$ if and only if $W \preccurlyeq W^+$.
\end{definition}

The upgraded (degraded) approximation remains upgraded (degraded) during combining and splitting:

\newlength{\yuckkyhack}
\settowidth{\yuckkyhack}{$Z(W)$}
\begin{lem}[{\cite[Lem.~4.7]{Korada_PhD} \& \cite[Lem.~3]{Tal_PolarConstruction}}]\label{lem:grading}
 Assume that $W^- \preccurlyeq W\preccurlyeq W^+$. Then,
 \begin{subequations}
   \begin{alignat}{2}
    I(W^-) &\le \makebox[\yuckkyhack]{$I(W)$} & &\le I(W^+)\\
    Z(W^-) &\ge Z(W) & &\ge Z(W^+)\\
    (W^-)_2^1 &\preccurlyeq \makebox[\yuckkyhack]{$W_2^1$} & &\preccurlyeq (W^+)_2^1\\
    (W^-)_2^0 &\preccurlyeq \makebox[\yuckkyhack]{$W_2^0$} & &\preccurlyeq (W^+)_2^0.
  \end{alignat}
 \end{subequations}
\end{lem}

It can be shown that the better channel~\eqref{eq:better} obtained from combining and splitting is upgraded w.r.t.\ the original channel (as already mentioned in~\cite[p.~9]{Sasoglu_PolarCodes}). The worse channel~\eqref{eq:worse} is degraded at least if $W$ is symmetric.

\begin{lem}[{\cite[p.~9]{Sasoglu_PolarCodes}}~\&{~\cite[Lem.~3]{Bardet_DecreasingMonomial}}]\label{lem:betterisupgraded}
$W\preccurlyeq W_2^1$. If $W$ is symmetric, then $W_2^0  \preccurlyeq W\preccurlyeq W_2^1$.
\end{lem}

\begin{IEEEproof}
By choosing
 \begin{equation}
  P(y|y_1^2,u_1) =\begin{cases}
                     1,& \text{if } y=y_2\\ 0, & \text{else.}
                    \end{cases}
 \end{equation}
 one can show that $W\preccurlyeq W_2^1$. To show that also $W_2^0  \preccurlyeq W$ for symmetric channels, take~\cite[Lem.~3]{Bardet_DecreasingMonomial}
 \begin{equation}
  Q(y_1^2|y) = \begin{cases}
                \frac{1}{2}W(y_2|0) & \text{ if } y_1=y\\
                \frac{1}{2}W(y_2|1) & \text{ if } y_1=\pi(y)\\
                0 & \text{ else}
               \end{cases}.
 \end{equation}
\end{IEEEproof}

\begin{example}
 For a BEC $W$ with erasure probability $\epsilon$, $W_2^1$ is a BEC with erasure probability $\epsilon^2$ and $W_2^0$ is a BEC with erasure probability $2\epsilon-\epsilon^2$~\cite[Prop.~6]{Arikan_PolarCodes}. The channel $W_2^1$ is an upgrade of $W$, because it can be degraded to $W$ by appending a BEC with erasure probability $\epsilon/(1+\epsilon)$. The channel $W_2^0$ is degraded w.r.t.\ $W$ by appending a BEC with erasure probability $\epsilon$. 
\end{example}

\section{Properties of the Sets $\Gset$ and $\Bset$}\label{sec:polar}
In this section we develop the properties of the sets of good and bad channels.

\begin{prop}\label{prop:basins}
 For almost all $x$, there exists a value $0\le\vartheta(x)\le 1$ such that $Z(W)<\vartheta(x)$ implies $x\in\Gset$. If $W$ is a BEC, then additionally $Z(W)>\vartheta(x)$ implies $x\in\Bset$.
\end{prop}

\begin{IEEEproof}
See Appendix~\ref{proof:basins}.
\end{IEEEproof}

If $W$ is not a BEC, it may happen that $Z(W)>\vartheta(f(b))$ while still $I(W_\infty^b)=1$. This leads to the question whether the set of good channels is (almost surely) increasing with decreasing Bhattacharyya parameter, i.e., if the sets of good channels for $W$ and $W'$ with $Z(W)>Z(W')$ are \emph{aligned}. While in general the answer is negative~\cite{Renes_Aligned}, Proposition~\ref{prop:basins} answers it positively if $W$ is a BEC: The set of good channels for a BEC is also good for any binary-input memoryless channel with a smaller Bhattacharyya parameter~\cite{Hassani_Compound}.

\begin{example}
 For $x\in\mathbb{D}$, $\vartheta(x)=1$: If $Z(W)<1$, i.e., if the channel is not completely useless a priori, the non-terminating expansion of $x$ will make it a perfect channel (cf.~Proposition~\ref{prop:dyadic}).
\end{example}

In Appendix~\ref{proof:symmetry} we prove that the thresholds of Proposition~\ref{prop:basins} are symmetric:
\begin{prop}\label{prop:symmetry}
 For those $x\notin\mathbb{D}$ for which $\vartheta(x)$ exists, $\vartheta(1-x) = 1-\vartheta(x)$.
\end{prop}

The case $x\in\mathbb{Q}\setminus\mathbb{D}$ is interesting. In this case, the binary expansion is unique and \emph{recurring}, i.e., there is a length-$k$ sequence $a^k\in\{0,1\}^k$, such that $f(b^na^ka^ka^k\cdots)=x$ for some $b^n\in\{0,1\}^n$. It is straightforward to show that for every non-trivial sequence $a_k$ (i.e., $a_k$ contains zeros and ones), $p_{a^k}$ is from $[0,1]$ to $[0,1]$, non-negative, and non-decreasing, with vanishing derivatives at 0 and 1. Since this ensures that $p_{a^k}(z)<z$ for $z$ close to zero and $p_{a^k}(z)>z$ for $z$ close to one, the operation $z_{i+1}=p_{a^k}(z_i)$ constitutes an iterated function system with attracting fixed points at $z=0$ and $z=1$. Note further that, since $p_{a^k}$ corresponds to the recurring part of the binary expansion of $x$, $Z(W_\infty^{b^na^ka^k\cdots})$ will be bounded from above by the value to which this iterated function system converges after being initialized with $Z(W_{2^n}^{b^n})$. To show that Proposition~\ref{prop:basins} holds for $x\in\mathbb{Q}\setminus\mathbb{D}$ requires showing that $p_{a^k}$ intersects the identity function only once on $(0,1)$, i.e., that there is no attracting fixed point on this open interval. We leave this problem for future investigation.

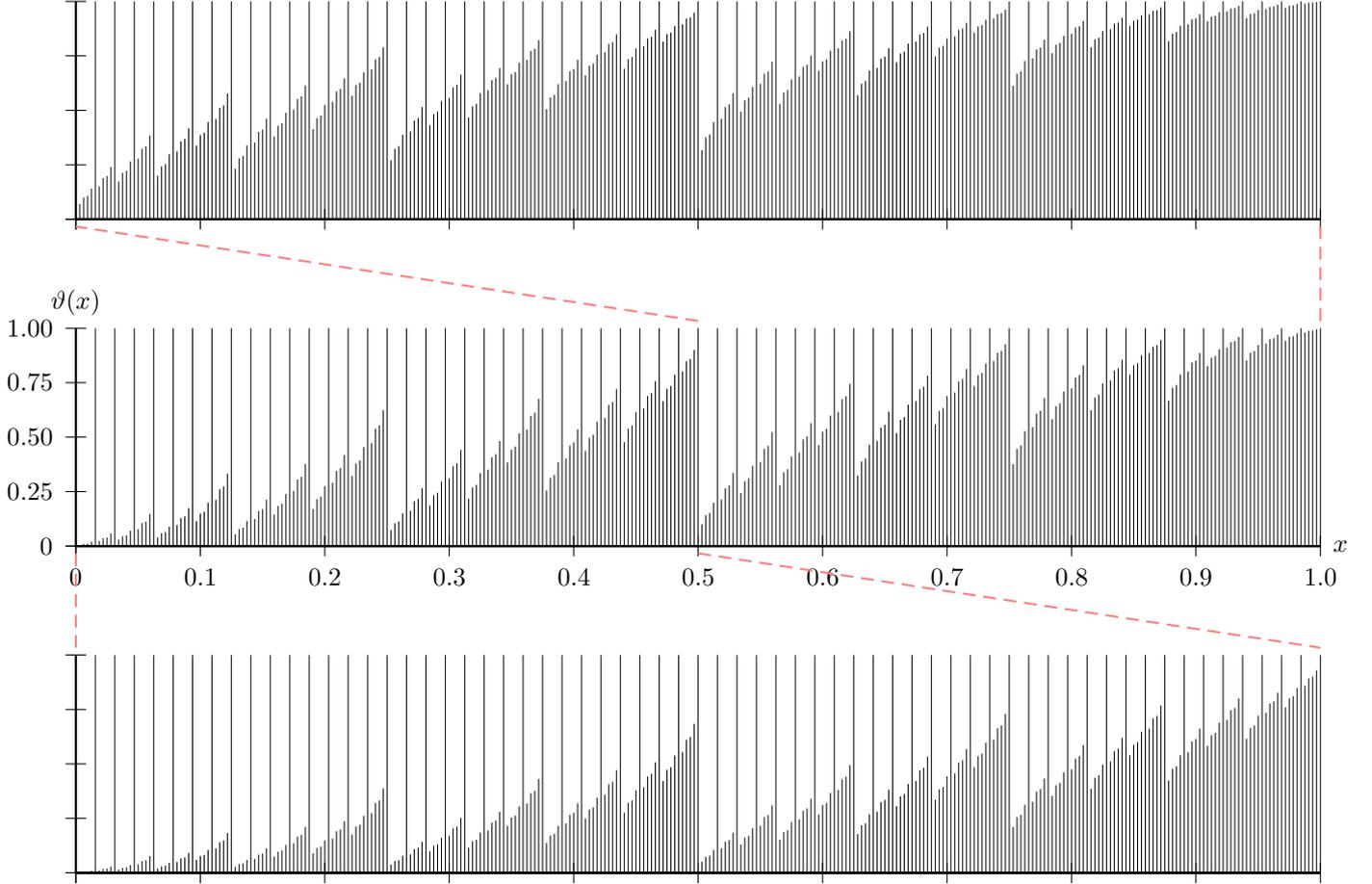
\begin{figure*}
 \begin{pspicture}[showgrid=false](0,-5)(18,7.5)
   \rput(0.5,4.5){
  \psaxes[xLabels={\empty},yLabels={\empty},Dx=0.05,dx=1.7,Dy=0.25,dy=0.75,Ox=0.5]{-}(0,0)(17,3)[,-90][,0]
  \readdata{\foo}{Base.dat}
  \psset{xunit=34cm,yunit=3cm}
  \dataplot[plotstyle=LineToXAxis,linewidth=0.1pt,xStart=0.5,origin={-0.5,0}]{\foo}
  }
  \rput(0.5,0){
  \psaxes[Dx=0.1,dx=1.7,Dy=0.25,dy=0.75]{-}(0,0)(17,3)[$x$,0][$\vartheta(x)$,90]
  \readdata{\foo}{Base.dat}
  \psset{xunit=17cm,yunit=3cm}
  \dataplot[plotstyle=LineToXAxis,linewidth=0.1pt,nStep=2,nStart=1]{\foo}
  }
  \rput(0.5,-4.5){
  \psaxes[xLabels={\empty},yLabels={\empty},Dx=0.05,dx=1.7,Dy=0.25,dy=0.75]{-}(0,0)(17,3)[,-90][,90]
  \readdata{\foo}{Base.dat}
  \psset{xunit=34cm,yunit=3cm}
  \dataplot[plotstyle=LineToXAxis,linewidth=0.1pt,xEnd=0.5]{\foo}
  }
  \psline[linecolor=red!50,linestyle=dashed](0.5,-.1)(0.5,-1.4)
  \psline[linecolor=red!50,linestyle=dashed](9,-.1)(17.5,-01.4)
  
  \psline[linecolor=red!50,linestyle=dashed](17.5,3.1)(17.5,4.4)
  \psline[linecolor=red!50,linestyle=dashed](9,3.1)(0.5,4.4)
 \end{pspicture}
 \caption{The polar fractal for a BEC. The center plot shows the thresholds $\vartheta(x)$ for $x\in[0,1]$, while the bottom and the top plots show these thresholds for the scaled and shifted sets $[0,0.5]$ and $[0.5,1]$, respectively. Hence, the thresholds in the top plot are larger than the thresholds in the center plot, which are larger than those in the bottom plot. The indicator function of $\Gset$ is obtained by setting each value in the plot to one (zero) if the erasure probability $\epsilon$ is smaller (larger) than the threshold. Note further that the figure illustrates the symmetry of $\vartheta(x)$ mentioned in Proposition~\ref{prop:symmetry}.}
 \label{fig:polarfractal}
\end{figure*}

\begin{example}\label{ex:goldenratio}
 Let $x=2/3$, hence $f^{-1}(x)=101010101\cdots$. It suffices to consider one period of the recurring sequence and determine its fixed points. In this case we get $p_{10}(z)=2z^2-z^4$. Its fixed points are the roots of $p_{10}(z)-z$; removing the trivial roots at $z=0$ and $z=1$ leaves two further roots at $(\pm\sqrt{5}-1)/2$. One of these roots lies outside $[0,1]$ and is hence irrelevant. The remaining root determines the threshold, $\vartheta(2/3)=(\sqrt{5}-1)/2$. 
 
 Let $W$ be a BEC with erasure probability $\epsilon=Z(W)=\vartheta(2/3)$. Since $\epsilon=\vartheta(2/3)$ is a fixed point of the iterated function system corresponding to the recurring binary expansion, one gets $Z(W_\infty^{f^{-1}(2/3)})=\epsilon\notin\{0,1\}$. This example illustrates that Proposition~\ref{prop:polarization} holds only almost surely.
\end{example}

\begin{prop}\label{prop:dyadic}
$\Gset\cap\Bset = \mathbb{D}$.
\end{prop}

\begin{IEEEproof}
See Appendix~\ref{proof:dyadic}.
\end{IEEEproof}

That the intersection of the sets of good and bad channels is non-empty is a direct consequence of the non-injectivity of $f$. Note further that this intersection cannot be larger, since $\mathbb{D}$ is the only set to which $f$ maps non-injectively. Since $\mathbb{D}$, a common subset of $\Gset$ and $\Bset$, is dense in $[0,1]$, both the set of good channels and the set of bad channels are dense in the unit interval. But even if dyadic rationals are excluded, results about denseness can be proved:

\begin{prop}\label{prop:dense}
 $\Gset\setminus\mathbb{D}$ is dense in $[0,1]$. If $W$ is a BEC, then also $\Bset\setminus\mathbb{D}$ is dense in $[0,1]$.
\end{prop}

\begin{IEEEproof}
See Appendix~\ref{proof:dense}.
\end{IEEEproof}

The proposition states that, at least for the BEC, there is no interval which contains only good channels. Hence, given a specific channel $W_{2^n}^{b^n}$, it is not possible to assume that a well-specified subset of channels (e.g., all $W_{\infty}^{b^na}$ for $a$ starting with $1$) generated from this channel by combining and splitting will be perfect. The construction algorithm for an infinite-blocklength, \emph{vanishing-error} polar code hence cannot stop at a finite blocklength. This is in contrast with finite-blocklength polar codes, for which an approximate construction technique suggests to stop polarizing some channels at already shorter blocklengths~\cite{ElKhamy_Relaxed}.

\begin{prop}\label{prop:lebesgue}
 $\Gset$ is Lebesgue measurable and has Lebesgue measure $\lambda(\Gset)=I(W)$. $\Bset$ is Lebesgue measurable and has Lebesgue measure $\lambda(\Bset)=1-I(W)$.
\end{prop}

\begin{IEEEproof}
See Appendix~\ref{proof:lebesgue}.
\end{IEEEproof}

Note that $\lambda(\Gset\cup\Bset)=1$ although $\Gset\cup\Bset\subset[0,1]$. The reason is that convergence to good or bad channels is only almost sure, i.e., there may be channels $W_\infty^b$ that are neither good nor bad (see Example~\ref{ex:goldenratio}).

An immediate consequence of Proposition~\ref{prop:lebesgue} is that $\Gset$ and $\Bset$ have a Hausdorff dimension equal to one. This follows from the fact that the one-dimensional Hausdorff measure of a set equals its Lebesgue measure up to a constant~\cite[eq.~(3.4), p.~45]{Falconer_Fractals}. Since, thus, the one-dimensional Hausdorff measures of $\Gset$ and $\Bset$ are positive and finite, we have
\begin{cor}\label{cor:hausdorff}
 The Hausdorff dimensions of $\Gset$ and $\Bset$ satisfy $\infodim{\Gset}=1$ and  $\infodim{\Bset}=1$.
\end{cor}
Also the box-counting dimensions~\cite[p.~28]{Falconer_Fractals} are equal to one, since both sets are dense on the unit interval~\cite[Prop.~2.6]{Falconer_Fractals}.

We finally come to the claim that polar codes are fractal. Following Falconer's definition~\cite[p.~xxviii]{Falconer_Fractals}, a set is fractal if it is (at least approximately) self-similar and has detail on arbitrarily small scales, or if its fractal dimension (e.g., its Hausdorff dimension) is larger than its topological dimension. Whether or not the result shown below will convince the reader of this property is a mere question of definition; strictly speaking, we can show only \emph{quasi self-similarity} of $\Gset$:

\begin{prop}\label{prop:selfsimilar}
 Let $\Gset_n(k):=\Gset \cap [(k-1)2^{-n},k2^{-n}]$ for $k=1,\dotsc,2^n$. $\Gset=\Gset_0(1)$ is \emph{quasi self-similar} in the sense that, for all $n$ and all $k$, $\Gset_n(k)={\Gset}_{n+1}(2k-1)\cup{\Gset}_{n+1}(2k)$ is quasi self-similar to its right half:
 \begin{equation}
  \Gset_{n}(k)\subset2{\Gset}_{n+1}(2k)-k2^{-n}
 \end{equation}
 If $W$ is symmetric, $\Gset_n(k)$ is quasi self-similar:
  \begin{equation}
 2{\Gset}_{n+1}(2k-1)-(k-1)2^{-n} \subset \Gset_{n}(k)\subset 2{\Gset}_{n+1}(2k)-k2^{-n}
 \end{equation}
\end{prop}

\begin{IEEEproof}
See Appendix~\ref{proof:selfsimilar}.
\end{IEEEproof}

In other words, at least for a symmetric channel, $\Gset$ is composed of two similar copies of itself (see Fig.~\ref{fig:polarfractal}). The self-similarity is closely related to the fact that polar codes are decreasing monomial codes~\cite[Thm.~1]{Bardet_DecreasingMonomial}. Along the same lines, the quasi self-similarity of $\Bset$ can be shown.

\begin{example}
 By careful computations we obtain $\vartheta(1/6)\approx 0.214$, $\vartheta(1/3)\approx 0.382$, and $\vartheta(2/3)\approx 0.618$. Indeed, if we consider $1/3$ in $\Gset$, then $1/6$ and $2/3$ are the corresponding values in $\Gset_1(1)$ and $\Gset_1(2)$. Since $\vartheta(1/6)<\vartheta(1/3)<\vartheta(2/3)$, for the BEC we have the inclusion indicated in Proposition~\ref{prop:selfsimilar}.
\end{example}

\section{Preliminaries for Reed-Muller Codes}\label{sec:prelim:RM}
As mentioned above, a rate-$K/2^n$ Reed-Muller code has a $K\times 2^n$ generator matrix with all $K$ rows having a Hamming weight larger than a predefined threshold. To make this more precise, let $w(b^n)=\sum_{i=1}^n b_i$ be the \emph{Hamming weight} of $b^n\in\{0,1\}^n$ and let $s_i(n)$ be the $i$-th row of $G(n)$. The generator matrix $G_{RM}(r,n)$ of an order-$r$, length-$2^n$ Reed-Muller code consists of the rows of $G(n)$ indicated in~\cite{Arikan_PolarRM}
\begin{equation}
 \Fset =\{i\in\{1,\dotsc,2^n\}{:}\ w(s_i(n))\ge 2^{n-r} \}.
\end{equation}
Trivially, $G_{RM}(n,n)=G(n)$, while $G_{RM}(0,n)$ is a single row vector containing only ones (length-$2^n$ repetition code).

To analyze the effect of doubling the block length, note that
\begin{equation}
 G(n+1):=\left[\begin{array}{cc}
           G(n) & 0\\ G(n) & G(n)
          \end{array}
\right].\label{eq:RMiteration}
\end{equation}
Assume that we indicate the rows of $G(n)$ by a sequence of binary numbers, i.e., let the $i$-th row be indexed by $h_n(b^n):=2^n\sum_{l=1}^n b_l 2^{-l}$. Furthermore, let $0b^n$ and $1b^n$ denote the sequences of zeros and ones obtained be prepending 0 and 1 to $b^n$, respectively. Clearly, $h_{n+1}(0b^n)=h_n(b^n)$ and $h_{n+1}(1b^n)=h_n(b^n)+2^n$. Combining this with~\eqref{eq:RMiteration} yields
\begin{align}
 w(s_{h_{n+1}(0b^n)}(n+1))&=w(s_{h_n(b^n)}(n))\\
 w(s_{h_{n+1}(1b^n)}(n+1))&=2w(s_{h_n(b^n)}(n)).
\end{align}
Defining $G(0):=1$, we thus get
\begin{equation}
 w(s_{h_n(b^n)}(n)) = 2^{w(b^n)}\label{eq:powerofweight}
\end{equation}
and
\begin{equation}
 \Fset =h_n\left(\{b^n\in\{0,1\}^n{:}\ 2^{w(b^n)} \ge 2^{n-r} \}\right).
\end{equation}

Letting the blocklengths go to infinity, we may ask questions about the following set:
\begin{definition}[The Heavy Channels]\label{def:heavy}
Let $\Hset(\rho)$ denote the set of $\rho$-heavy channels, i.e.,
\begin{equation}
 x\in \Hset(\rho) \Leftrightarrow \exists b\in f^{-1}(x){:}\ \liminf_{n\to\infty} \frac{2^{w(b^n)}}{2^{n\rho}} \ge 1.
\end{equation}
\end{definition}
Loosely speaking, the set of heavy channels corresponds to those rows of $G(n)$, which \emph{asymptotically} have a Hamming weight larger than a given threshold.

\begin{example}\label{ex:heavyH}
 $\Hset(1)=\{1\}$. This follows from the fact that 1 is the only number in the unit interval with a binary expansion consisting only of ones. $\Hset(0)=[0,1]$. This follows from the fact that $w(b^n)\ge 0$. 
\end{example}

The results we will show for the set $\Hset(\rho)$ are tightly linked to the concept of \emph{normal numbers}.
\begin{definition}[Normal Numbers]\label{def:normal}
 A number $x\in[0,1]$ is called \emph{simply normal to base 2} ($x\in\mathcal{N}$) iff
 \begin{equation}
  \exists b\in f^{-1}(x){:}\ \limn\frac{w(b^n)}{n} = \frac{1}{2}.
 \end{equation}
\end{definition}

In general, a number is simply normal in base $M$ if the number of each of its digits used in its $M$-ary expansion is $1/M$. A number is called normal if this property not only holds for digits, but for subsequences: a number is normal in base $M$ if, for each $k\ge 1$, the number of each of its length-$k$ sequences used in its $M$-ary expansion is $1/M^k$. It immediately follows that a normal number is simply normal. The converse is in general not true:

\begin{example}\label{ex:normal}
 Let $x=1/3$, hence $b=010101\cdots$. $x$ is simply normal to base 2, but not normal (since the sequences $00$ and $11$ never occur). Let $x=1/7$, hence $b=001001001\cdots$. $x$ is neither normal nor simply normal. Let $x\in\mathbb{D}$, hence $b$ is either terminating ($\limn w(b^n)/n=0$) or non-terminating ($\limn w(b^n)/n=1$). Dyadic rationals are not simply normal.
\end{example}

\begin{lem}[Borel's Law of Large Numbers, cf.~{\cite[Cor.~8.1, p.~70]{Kuipers_Uniform}}]\label{lem:borel}
 Almost all numbers in $[0,1]$ are simply normal, i.e.,
 \begin{equation}
  \lambda(\mathcal{N})=1.
 \end{equation}
\end{lem}

Although normal numbers are, in this sense, \emph{normal}, there are uncountably many numbers in the unit interval which are not normal. Moreover, the set of numbers that are not normal is \emph{superfractal}, i.e., it has a Hausdorff dimension equal to one although it has zero Lebesgue measure~\cite{Albeverio_Superfractal}.

\section{Properties of the Set $\Hset$}\label{sec:reedmuller}
We can show in Appendix~\ref{proof:dyadic:RM} that the dyadic rationals are not only good and bad, but also heavy:
\begin{prop}\label{prop:dyadic:RM}
For all $\rho\in[0,1)$, $\mathbb{D}\subset \Hset(\rho)$. 
\end{prop}

It follows that $\Hset(\rho)$ is dense in $[0,1]$ for all $\rho\in[0,1)$.

The Lebesgue measure of the set of good channels was equal to the channel capacity of $W$. The result for heavy channels is inherently different, because $\Hset(\rho)$ does not depend on $W$. The proof of the following result can be found in Appendix~\ref{proof:lebesgue:RM}.
\begin{prop}\label{prop:lebesgue:RM}
 $\Hset(\rho)$ is Lebesgue measurable and has Lebesgue measure
 	\begin{equation}
        \lambda(\Hset(\rho))=\begin{cases}
                                   1, & \text{ if } \rho<1/2\\ 0, & \text{ if } \rho\ge1/2
                                  \end{cases}.
       \end{equation}
\end{prop}

The result is surprising since it suggests a \emph{phase transition} for the rate of Reed-Muller codes: If $\rho<1/2$, the infinite-blocklength Reed-Muller code consists of almost all (in the sense of Lebesgue measure) possible binary sequences. In contrast, if $\rho\ge 1/2$, the infinite-blocklength Reed-Muller code consists of almost no code words (again, in the sense of Lebesgue measure). The picture is not as simple if one also considers the Hausdorff dimension of $\Hset(\rho)$. In Appendix~\ref{proof:hausdorff:RM} we prove that $\Hset(\rho)$ has positive Hausdorff dimension even if it is a Lebesgue null set.
\begin{prop}\label{prop:hausdorff:RM}
 The Hausdorff dimension satisfies 
	\begin{equation}
        \infodim{\Hset(\rho)}\begin{cases}
                                   =1, & \text{ if } \rho\le 1/2\\ \ge h_2(\rho), & \text{ if } \rho>1/2
                                  \end{cases}
       \end{equation}
       where $h_2(x):=-x\log_2 x - (1-x)\log_2 (1-x)$.
\end{prop}

Unfortunately, we were not able to give an exact expression for the Hausdorff dimension of $\Hset(\rho)$ for $\rho>1/2$. While the set of all non-normal numbers is superfractal, we are not sure if this holds also for a proper subset.

The sets $\Gset$ and $\Bset$ exhibit self-similarity, i.e., detailed structure on every scale (cf. Fig.~\ref{fig:polarfractal}). We next show that also $\Hset(\rho)$ is self-similar. At least for $\Hset(0)$ and $\Hset(1)$ (cf.~Example~\ref{ex:heavyH}) this is as trivial as the self-similarity of a point or a line. For $\rho\in(0,1)$ this self-similarity is more interesting, and related to the fact that Reed-Muller codes are decreasing monomial codes~\cite[Prop.~2]{Bardet_DecreasingMonomial}. In Appendix~\ref{proof:selfsimilar:RM} we prove
\begin{prop}\label{prop:selfsimilar:RM}
Let $\Hset_n(\rho,k):=\Hset(\rho) \cap [(k-1)2^{-n},k2^{-n}]$ for $k=1,\dotsc,2^n$. $\Hset(\rho)=\Hset_0(\rho,1)$ is \emph{quasi self-similar} in the sense that, for all $n$ and all $k$, $\Hset_n(\rho,k)={\Hset}_{n+1}(\rho,2k-1)\cup{\Hset}_{n+1}(\rho,2k)$ is quasi self-similar:
 \begin{multline}
 2{\Hset}_{n+1}(\rho,2k-1)-(k-1)2^{-n} \subset\\ \Hset_{n}(\rho,k)\subset 2{\Hset}_{n+1}(\rho,2k)-k2^{-n}.
 \end{multline}
\end{prop}

\section{Discussion \& Outlook}\label{sec:discussion}
That polar codes satisfy fractal properties has long been suspected: Every nontrivial, partly polarized channel $W_{2^n}^{b^n}$ gives rise, by further polarization, to both perfect and useless channels, regardless how close $I(W_{2^n}^{b^n})$ is to zero or one. This fact is reflected in our Propositions~\ref{prop:dyadic} and~\ref{prop:dense}, which state that the good channels are dense in the unit interval (and so are the bad channels for BECs): A partial polarization with sequence $b^n$ corresponds to an interval with dyadic endpoints, and denseness implies that in this interval there will be both perfect and useless channels. Proposition~\ref{prop:selfsimilar}, claiming the self-similarity of the sets of good and bad channels, goes one step further and gives these sets structure: If a channel polarized according to the sequence $b^na$ is good, then so is the channel polarized according to $b^n1a$. 

An obvious extension of our work should deal with the fractal properties of non-binary polar and Reed-Muller codes. For example, if $q$ is a prime number, then every invertible $\ell\times\ell$ matrix with entries from $\{0,\dots,q-1\}$ is polarizing, unless it is upper-triangular~\cite[Thm.~5.2]{Sasoglu_PolarCodes}. The $n$-fold Kronecker product of one of these matrices generates $\ell^n$ channels. It is easy to design a function mapping $\{0,\dots,\ell-1\}^\infty$ to $[0,1]$ (cf.~\eqref{eq:functionf}), admitting an analysis similar to the one presented in this paper. Along the same lines, it would be interesting to examine the properties of $q$-ary Reed-Muller codes, e.g.,~\cite{Kasami_RMCodes,Delsarte_RMCodes}.

Whether binary or not, it is presently not clear how our infinite-blocklength results can be carried over to practically relevant finite-length codes. Future work shall investigate this issue.

\section*{Acknowledgments}
The author thanks Emmanuel Abbe, Princeton University, and Hamed Hassani, ETH Zurich, for fruitful discussions and suggesting material. The author is particularly indebted to Jean-Pierre Tillich, INRIA, for helpful suggestions and generalizing Proposition~\ref{prop:selfsimilar}.

\appendices
\section{Proof of Proposition~\ref{prop:basins}}\label{proof:basins}
Recall that, by Lemma~\ref{lem:Bhattacharyya}, we have
\begin{equation*}
 Z(W_{2^n}^{b^n}) \le p_{b^n}(Z(W)):=g_{b_n}(g_{b_{n-1}}(\cdots g_{b_1}(Z(W)) \cdots )).
\end{equation*}
\begin{lem}[{\cite[Lem.~11]{Hassani_FiniteLength}}]\label{lem:thresholds}
 For $\Prob$-almost every realization $b\in\Omega$, there exists a point $\theta(b)\in[0,1]$, such that
 \begin{equation}
  \limn p_{b^n}(z)=\begin{cases}
                0, & z\in[0,\theta(b))\\ 1, & z\in(\theta(b),1]
               \end{cases}.
 \end{equation}
 Furthermore, the thus constructed RV $\theta$ is uniformly distributed on $[0,1]$.
\end{lem}

If $Z(W)<\theta(b)$, $Z(W_\infty^b)\le \limn p_{b^n}(Z(W))=0$, and hence $f(b)\in\Gset$. We now define $\vartheta(f(b)):=\theta(b)$ if $f(b)\notin\mathbb{D}$ and $\vartheta(f(b))=1$ if $f(b)\in\mathbb{D}$ (since $\mathbb{D}\subset\Gset$ by Proposition~\ref{prop:dyadic}).

\begin{IEEEproof}[Proof for BECs]
 If $W$ is a BEC, then $Z(W_{2^n}^{b^n}) = p_{b^n}(Z(W))$. Hence, by Lemma~\ref{lem:thresholds}, if $\epsilon<\theta(b)$, then $Z(W_\infty^b)=\limn p_{b^n}(\epsilon)=0$, and if $\epsilon>\theta(b)$, then $Z(W_\infty^b)=\limn p_{b^n}(\epsilon)=1$.
\end{IEEEproof}

\section{Proof of Proposition~\ref{prop:symmetry}}\label{proof:symmetry}
Let $b\in f^{-1}\left([0,1]\setminus\mathbb{D}\right)\subset\{0,1\}^\infty$, and let $\overline{b}$ be such that $\overline{b}_i=1-b_i$ for all $i$. It follows from the linearity of $f$ that $f(b)+f(\overline{b})=f(b+\overline{b})=1$, because $b+\overline{b}=11111\cdots$. Hence, if $x\notin\mathbb{D}$ has binary expansion $b$, then $1-x$ has binary expansion $\overline{b}$. 
It can be easily verified that $g_i(1-z)=1-g_{1-i}(z)$ for $i=0,1$. Hence,
\begin{align*}
 p_{b^n}(z)&=g_{b_n}(g_{b_{n-1}}(\cdots g_{b_2}(g_{b_1}(z)) \cdots ))\\
 &= g_{b_n}(g_{b_{n-1}}(\cdots g_{b_2}(1-g_{\overline{b}_1}(1-z)) \cdots ))\\
 &= g_{b_n}(g_{b_{n-1}}(\cdots 1-g_{\overline{b}_2}(g_{\overline{b}_1}(1-z)) \cdots ))\\
 &= 1-g_{\overline{b}_n}(g_{\overline{b}_{n-1}}(\cdots g_{\overline{b}_2}(g_{\overline{b}_1}(1-z)) \cdots ))\\
 &= 1-p_{\overline{b}^n}(1-z).
\end{align*}
If $0\le z<\theta(b)$, then $1-\theta(b)<1-z\le 1$. Since $0\le z<\theta(b)$ implies $p_{b^n}(z)\to0$ and $p_{\overline{b}^n}(1-z)\to1$, we get $\theta(\overline{b})=1-\theta(b)$, and hence $\vartheta(1-x)=1-\vartheta(x)$. This completes the proof.\hfill\IEEEQED

\section{Proof of Proposition~\ref{prop:dyadic}}\label{proof:dyadic}
That $\Gset\cap\Bset\subseteq\mathbb{D}$ follows from the fact that only dyadic rationals have a non-unique binary expansion. In particular, the preimage of every $x\in\mathbb{D}$ consists of two elements, namely
\begin{subequations}
 \begin{equation}
 (b^{n-1}b_n 0000000\cdots) \label{eq:dyadic:worse}
\end{equation}
and
\begin{equation}
(b^{n-1}\overline{b}_n 1111111\cdots) \label{eq:dyadic:better}
\end{equation}
\end{subequations}
where $\overline{b}_n=1-b_n$. By the properties of combining and splitting,
\begin{equation}
 0<I(W_{2^n}^{b^{n-1}\overline{b}_n}), I(W_{2^n}^{b^{n-1}b_n}) <1.
\end{equation}
We first show that~\eqref{eq:dyadic:better} leads to a good channel. To this end, observe that, by~\cite[Prop.~7]{Arikan_PolarCodes}, the Bhattacharyya parameter satisfies $0<Z(W_{2^{n+1}}^{b^{n-1}\overline{b}_n1})=Z(W_{2^{n}}^{b^{n-1}\overline{b}_n})^2<1$. Iterating the squaring operation drives the Bhattacharyya parameter to zero, i.e., $Z(W_\infty^{b^{n-1}\overline{b}_n 1111111\cdots}) = 0$, hence $I(W_\infty^{b^{n-1}\overline{b}_n 1111111\cdots})=1$ and $\mathbb{D}\in\Gset$.

To show that~\eqref{eq:dyadic:worse} leads to a bad channel, assume that $I(W_{\infty}^{b^n0000000\cdots})=\delta$. We now show that for every $a\in\Omega$, $I(W_\infty^{b^n0000000\cdots})\le I(W_\infty^{b^na})$. For example, take $a=(1011100\cdots)$. By Lemmas~\ref{lem:grading} $(a)$ and~\ref{lem:betterisupgraded} $(b)$, the following list of relations can be shown:
\begin{align*}
 W_\infty^{b^n} &\stackrel{(a)}{\preccurlyeq} W_\infty^{b^n1}\\
 W_\infty^{b^n0} &\stackrel{(b)}{\preccurlyeq} W_\infty^{b^n10}\\
 W_\infty^{b^n0} &\stackrel{(a)}{\preccurlyeq} W_\infty^{b^n101}\\
 W_\infty^{b^n0} &\stackrel{(a)}{\preccurlyeq} W_\infty^{b^n1011}\\
 W_\infty^{b^n0} &\stackrel{(a)}{\preccurlyeq} W_\infty^{b^n10111}\\
 W_\infty^{b^n00} &\stackrel{(b)}{\preccurlyeq} W_\infty^{b^n101110}\\
 W_\infty^{b^n000} &\stackrel{(b)}{\preccurlyeq} W_\infty^{b^n1011100}\\
 \dots &\preccurlyeq \dots
\end{align*}
and hence, $W_\infty^{b^n0000000\cdots} \preccurlyeq W_\infty^{b^na}$. By Lemma~\ref{lem:grading}, $\delta=I(W_\infty^{b^n0000000\cdots})\le I(W_\infty^{b^na})$ for every $a\in\Omega$, hence also
\begin{equation}
 \delta \le \inf_{a\in\Omega} I(W_\infty^{b^na}).
\end{equation}
But since $0<I(W_{2^n}^{b^n})<1$, by Proposition~\ref{prop:polarization} there must be sequences $a$ such that $I(W_\infty^{b^na})=0$, hence $\delta=0$ and $\mathbb{D}\in\Bset$.\hfill\IEEEQED

\section{Proof of Proposition~\ref{prop:dense}}\label{proof:dense}
The proof follows from showing that between every dyadic rational we can find a rational $x\in\mathbb{Q}\setminus\mathbb{D}$ such that $x\in\Gset$. To this end, fix $x_1=p/2^n$ and $x_2=(p+1)/2^n$. Let further $b^n$ be the terminating binary expansion of $x_1$, i.e., $f(b^n000\cdots)=x_1$. Let $a^k$ be such that $a_1=\cdots=a_{k-1}=1$ and $a_k=0$. Note that $x:=f(b^na^ka^ka^k\cdots)\in(x_1,x_2)$. We now bound the polynomial $p_{a^k}$ from above:
\begin{equation*}
 p_{a^k}(z)= 2z^{2^{k-1}}-z^{2^{k}}\le 2 z^{2^{k-1}}
\end{equation*}
The bound crosses $z$ at $z=0$ and at $z^*=2^{-1/(2^{k-1}-1)}$. From this follows that $p_{a^k}(z)<z$ for $z<z^*$, where $z^*$ can be made arbitrarily close to one for $k$ sufficiently large. Hence, if $z_{i+1}=p_{a^k}(z_i)$, then $z_i\to 0$ if $z_0<z^*$. Let $z_0=Z(W_{2^n}^{b_n})$ and let $k$ be sufficiently large such that $z^*>z_0$. Then, $Z(W_\infty^{b^na^ka^k\cdots})=0$ and $x\in\Gset$. 

\begin{IEEEproof}[Proof for BECs]
 It remains to show that also $\Bset\setminus\mathbb{D}$ is dense in $[0,1]$. To this end, we consider the sequence $a^k$ such that $a_1=\cdots=a_{k-1}=0$ and $a_k=1$. We now bound the polynomial $p_{a^k}$ from below:
 \begin{align*}
  p_{a^k}(z) &= \left(1-(1-z)^{2^{k-1}}\right)^2\\
  &= 1-2(1-z)^{2^{k-1}}+(1-z)^{2^{2k-2}}\\
  &\ge 1-2(1-z)^{2^{k-1}}
 \end{align*}
The bound crosses $z$ at at $z=1$ and $z^*=1-2^{-1/(2^{k-1}-1)}$. From this follows that $p_{a^k}(z)>z$ for $z>z^*$, where $z^*$ can be made arbitrarily close to zero for $k$ sufficiently large. Hence, if $z_{i+1}=p_{a^k}(z_i)$, then $z_i\to 1$ if $z_0>z^*$. Let $z_0=Z(W_{2^n}^{b_n})$ and let $k$ be sufficiently large such that $z^*<z_0$. Then, $Z(W_\infty^{b^na^ka^k\cdots})=1$ and $x\in\Bset$.
\end{IEEEproof}

\section{Proof of Proposition~\ref{prop:lebesgue}}\label{proof:lebesgue}
In the proof we use Lemma~\ref{lem:fprops}. Note that $\lambda(\Gset)=\lambda(\Gset\setminus\mathbb{D}) + \lambda(\mathbb{D}) = \lambda(\Gset\setminus\mathbb{D})$, and similarly, $\lambda(\Bset)=\lambda(\Bset\setminus\mathbb{D})$. Since $f$ is bijective on $\Omega':=\Omega\setminus f^{-1}(\mathbb{D})$, Definition~\ref{def:gset} implies
\begin{equation}
 x\in\Gset\setminus\mathbb{D} \Leftrightarrow I(W_\infty^{f^{-1}(x)})=1
\end{equation}
or $f^{-1}(\Gset\setminus\mathbb{D}) = \{b\in\Omega'{:}\ I(W_\infty^{b})=1\}$. Note further that $\Prob(f^{-1}(\mathbb{D}))=0$. Hence, by Proposition~\ref{prop:polarization},
\begin{multline}
 \lambda(\Gset)=\lambda(\Gset\setminus\mathbb{D}) = \Prob(\{b\in\Omega'{:}\ I(W_\infty^{b})=1\}) \\= \Prob(I_\infty=1)=I(W).
\end{multline}
The proof for the set of bad channels follows along the same lines.\hfill\IEEEQED

\section{Proof of Proposition~\ref{prop:selfsimilar}}\label{proof:selfsimilar}
Since the dyadic rationals are self-similar and since, by Proposition~\ref{prop:dyadic}, $\mathbb{D}\subset\Gset$, one has, for all $n$ and $k$,
\begin{equation}
 \Gset_n(k)\cap\mathbb{D} = 2\left(\Gset_{n+1}(2k)\cap\mathbb{D}\right)-k2^{-n}.
\end{equation}

We now treat those values in $[0,1]$ that are not dyadic rationals. If $b_k^n=b_1b_2\cdots b_n$ is the terminating binary expansion of $(k-1)2^{-n}$, every value in $[(k-1)2^{-n},k2^{-n}]$ has a binary expansion $b_k^na$ for some $a\in\Omega$, where $b_n=1$ if and only if $(k-1)$ is odd. Similarly, and since $(2k-1)$ is always odd, every value in $[(2k-1)2^{-n-1},k2^{-n}]$ has a binary expansion $b_k^n1a'$ for some $a'\in\Omega$. Assume that $a'=a$. Then, by Lemmas~\ref{lem:grading} and~\ref{lem:betterisupgraded}, $W_\infty^{b_k^na}\preccurlyeq W_\infty^{b_k^n1a}$ for all $a$. Hence, if $f(b_k^na)\in\Gset_n(k)$, then $f(b_k^n1a)\in\Gset_{n+1}(2k)$. It remains to show that $2f(b_k^n1a)-f(b_{k+1}^n)=f(b_k^na)$:
\begin{align*}
 f(b_k^na)+f(b_{k+1}^n) &= f(b_k^n)+2^{-n}f(a)+f(b_{k+1}^n)\\
 &= (k-1)2^{-n} + 2^{-n}f(a) + k2^{-n}\\
 &= (2k-1)2^{-n} + 2^{-n}f(a)\\
 &= 2(2k-1)2^{-n-1}+ 2\cdot2^{-n-1}f(a)\\
 &= 2f(b_k^n1)+2\cdot2^{-n-1}f(a)\\
 &= 2f(b_k^n1a)
\end{align*}

\begin{IEEEproof}[Proof for Symmetric Channels]
Since $(2k-2)$ is always even, every value in $[(2k-2)2^{-n-1},(2k-1)2^{-n-1}]$ has a binary expansion $b_k^n0a$ for some $a\in\Omega$. Then, by Lemmas~\ref{lem:grading} and~\ref{lem:betterisupgraded}, $W_\infty^{b_k^n0a}\preccurlyeq W_\infty^{b_k^na}$ for all $a$. Hence, if $f(b_k^n0a)\in\Gset_{n+1}(2k)$, then $f(b_k^na)\in\Gset_n(k)$. It remains to show that $2f(b_k^n0a)-f(b_{k}^n)=f(b_k^na)$:
\begin{align*}
 f(b_k^na)+f(b_{k}^n) &= f(b_k^n)+2^{-n}f(a)+f(b_{k}^n)\\
 &= (k-1)2^{-n} + 2^{-n}f(a) + (k-1)2^{-n}\\
 &= (2k-2)2^{-n} + 2^{-n}f(a)\\
 &= 2(2k-2)2^{-n-1}+ 2\cdot2^{-n-1}f(a)\\
 &= 2f(b_k^n0)+2\cdot2^{-n-1}f(a)\\
 &= 2f(b_k^n0a)
\end{align*}
\end{IEEEproof}

\section{Proof of Proposition~\ref{prop:dyadic:RM}}\label{proof:dyadic:RM}
 We take the non-terminating expansion of $x\in\mathbb{D}$, i.e., there is a $b^k\in\{0,1\}^k$ such that $f(b^k1111\cdots)=x$. Hence, $w(b^n)\ge n-k$ for $n\ge k$. In Definition~\ref{def:heavy} we can take the binary logarithm on both sides of the inequality to get the condition
 \begin{equation}
 x\in \Hset(\rho) \Leftrightarrow \exists b\in f^{-1}(x){:}\ \liminf_{n\to\infty}  w(b^n)-n\rho \ge 0.
\end{equation}
But
\begin{align}
 \liminf_{n\to\infty}  w(b^n)-n\rho  &= \limn n(1-\rho)-k
\end{align}
goes to infinity for $\rho<1$.\hfill\IEEEQED

\section{Proof of Proposition~\ref{prop:lebesgue:RM}}\label{proof:lebesgue:RM}
By Example~\ref{ex:normal}, dyadic rationals are not simply normal, hence let $\mathcal{N}\subset [0,1]\setminus\mathbb{D}$ be the set of simply normal numbers in $[0,1]$. Note that $f$ is bijective on $\mathcal{N}$ by Lemma~\ref{lem:fprops}. By Lemma~\ref{lem:borel} we have
\begin{equation}
 \forall b\in f^{-1}(\mathcal{N}){:}\ w(b^n)=\frac{1}{2}n+o(n).
\end{equation}
Fix $\rho$. Then,
\begin{equation}
 \liminf_{n\to\infty} w(b^n)-n\rho = \limn n\left(\frac{1}{2}-\rho\right) + o(n).
\end{equation}
If $\rho<1/2$, then this limit diverges to infinity, and hence $\mathcal{N}\subset \Hset(\rho)$. Thus, since $\lambda(\mathcal{N})=1$, we have $\lambda(\Hset(\rho))=1$. If $\rho>1/2$, the limit diverges to minus infinity, and hence $\mathcal{N}\not\subset \Hset(\rho)$. Thus, $\Hset(\rho)\subset [0,1]\setminus\mathcal{N}$, from which follows $\lambda(\Hset(\rho))= 0$.

Now let $\rho=1/2$. We define a random variable $B$ on our probability space, such that for all $b\in\Omega$, $B(b)=b$. $B$ is a sequence of independent, identically distributed Bernoulli-1/2 random variables, i.e., for all $i$ we have $\Prob(B_i=1)=\Prob(B_i=0)=1/2$. We have
\begin{equation}
 \lambda(\Hset(1/2)) = \Prob\left(\liminf_{n\to\infty} w(B^n)-\frac{n}{2}\ge 0\right).
\end{equation}
Consider the simple random walk $S_n:= w(B^n)-\frac{n}{2}$. Let $N_0(n)$ be the number of zero crossings of the sequence $S_1,\dotsc,S_n$, and let $N_0(n,b)$ be the number of zero crossings corresponding to the realization $b\in\Omega$. The event $\liminf_{n\to\infty} w(b^n)-\frac{n}{2}\ge 0$ can only happen if the realization of $S_n$ corresponding to $b$ has only finitely many zero crossings, i.e.,
\begin{align*}
 &\{b\in\Omega{:}\ \liminf_{n\to\infty} w(b^n)-\frac{n}{2}\ge 0\}\notag\\
 &\subseteq
 \{b\in\Omega{:}\ \exists R\in\mathbb{N}_0{:} \limn N_0(n,b)\le R\}\\
 &= \bigcup_{R=0}^\infty \{b\in\Omega{:}\ \limn N_0(n,b)\le R\}\\
 &= \bigcup_{R=0}^\infty \liminf_{n\to\infty} \{b\in\Omega{:}\ N_0(n,b)\le R\}
\end{align*}
and hence
\begin{multline}
 \Prob\left(\{b\in\Omega{:}\ \liminf_{n\to\infty} w(b^n)-\frac{n}{2}\ge 0\}\right)\\
 \le \sum_{R=0}^\infty \Prob(\liminf_{n\to\infty} \{b\in\Omega{:}\ N_0(n,b)\le R\})\\
 \le \sum_{R=0}^\infty \limn \Prob(N_0(n)\le R) \label{eq:sumOfProbabilities}
\end{multline}
where the second inequality is due to Fatou's lemma~\cite[Lem.~1.28, p.~23]{Rudin_Analysis3}.

With~\cite[Ch.~III.5,~p.~84]{Feller_Probability}
\begin{equation}
 \Prob(N_0(n)=R) = 2\Prob(S_{2n+1}=2R+1)
\end{equation}
we get
\begin{align*}
 \Prob(N_0(n)\le R)&=2\sum_{r=0}^R \Prob(S_{2n+1}=2r+1)\\
 &\stackrel{(a)}{=}2\sum_{r=0}^R {2n+1 \choose n-r} 2^{-2n-1}\\
 &\le 2^{-2n} \sum_{r=0}^R {2n+2 \choose n+1}\\
 &= 2^{-2n} (R+1) {2n+2 \choose n+1}\\
 &\stackrel{(b)}{\le} 2^{-2n} (R+1) e 2^{2n+2} \frac{1}{\sqrt{(n+1)\pi}}\\
 &= \frac{4e (R+1)}{\sqrt{(n+1)\pi}}
\end{align*}
where $(a)$ is~\cite[eq.~(2.2),~p.~75]{Feller_Probability} and $(b)$ is due to Stirling's approximation~\cite[eq.~6.1.38, p.~257]{Abramowitz_Handbook}. Since for $n\to\infty$ this probability is zero, we have by~\eqref{eq:sumOfProbabilities}
\begin{equation}
\lambda(\Hset(1/2)) = \Prob\left(\liminf_{n\to\infty} w(B^n)-\frac{n}{2}\ge 0\right) =0.
\end{equation}
This completes the proof.\hfill\IEEEQED


\section{Proof of Proposition~\ref{prop:hausdorff:RM}}\label{proof:hausdorff:RM}
That $\infodim{\Hset(\rho)}=1$ for $\rho<1/2$ follows from Proposition~\ref{prop:lebesgue:RM} in combination with Corollary~\ref{cor:hausdorff}.
For $\rho\ge 1/2$, we define
\begin{equation}
 \tilde{\mathcal{N}}_p := \left\{x\in[0,1]{:}\ \exists b\in f^{-1}(x){:} \limn \frac{w(b^n)}{n}=p\right\}
\end{equation}
for some $p\in (0,1)$. Note that $\tilde{\mathcal{N}}_{1/2}=\mathcal{N}$. By~\cite{Eggleston_FractalSets} (cf.~\cite[Chapter~8]{Kuipers_Uniform} for further notes), the Hausdorff dimension of this set is given by\footnote{Interestingly, in Eggleston's paper, the dimension was not connected to entropy; it was submitted earlier in the same year as Shannon's Mathematical Theory of Communication was published.}
\begin{equation}
 \infodim{\tilde{\mathcal{N}}_p}=h_2(p).
\end{equation}
Reasoning as in the proof of Proposition~\ref{prop:lebesgue:RM}, $\tilde{\mathcal{N}}_p\subset \Hset(\rho)$ if $p>\rho$ and $\tilde{\mathcal{N}}_p\not\subset \Hset(\rho)$ if $p<\rho$. As a consequence,
\begin{equation}
 \bigcup_{n=1}^\infty \tilde{\mathcal{N}}_{\rho+1/n}\subset\Hset(\rho).
\end{equation}
For a countable sequence of sets $A_n$, Hausdorff dimension satisfies~\cite[p.~49]{Falconer_Fractals}
\begin{equation}
 \infodim{\bigcup_{n=1}^\infty A_n} = \sup_{n\ge 1} \infodim{A_n},
\end{equation}
and hence, by the monotonicity of Hausdorff dimension~\cite[p.~48]{Falconer_Fractals},
\begin{equation}
 \infodim{\Hset(\rho)} \ge \sup_{n\ge 1} h_2(\rho+1/n) = h_2(\rho)
\end{equation}
where the last equality follows from the fact that the binary entropy function decreases with increasing $\rho$ for $\rho\ge 1/2$. In particular, for $\rho=1/2$, $\infodim{\Hset(\rho)} =1$. This completes the proof. \hfill\IEEEQED

\section{Proof of Proposition~\ref{prop:selfsimilar:RM}}\label{proof:selfsimilar:RM}
The proof follows along the lines of the proof of Proposition~\ref{prop:selfsimilar}. Let again $b_k^n$ be the terminating expansion of $(k-1)2^{-n}$, and let $a\in\Omega$. The connections between the sequences $b:=b_k^na$, $b_-:=b_k^n0a$, and $b_+:=b_k^n1a$ have been established above. To prove the theorem, we have to show that
\begin{subequations}
 \begin{align}
 & \liminf_{m\to\infty} w(b_-^m)-\rho m\ge 0 \label{eq:bminus}\\
 &\Rightarrow \liminf_{m\to\infty} w(b^m)-\rho m\ge 0\label{eq:b}\\
 &\Rightarrow \liminf_{m\to\infty} w(b_+^m)-\rho m\ge 0\label{eq:bplus}.
\end{align}
\end{subequations}
This is obtained by
\begin{subequations}
 \begin{align}
 &\liminf_{m\to\infty}w(b_-^m)-\rho m \notag\\
 &= w(b_k^n0)-\rho (n+1)  + \liminf_{m\to\infty}w(a^m)-\rho m\notag\\
 &= w(b_k^n)-\rho n-\rho  + \liminf_{m\to\infty}w(a^m)-\rho m\notag\\
 &\le w(b_k^n)-\rho n  + \liminf_{m\to\infty}w(a^m)-\rho m\label{eq:boundb}\\
 &\le w(b_k^n)-\rho n + (1-\rho)  + \liminf_{m\to\infty}w(a^m)-\rho m\notag\\
 &= w(b_k^n1)-\rho (n+1) + \liminf_{m\to\infty}w(a^m)-\rho m\label{eq:boundbplus}
\end{align}
\end{subequations}
where~\eqref{eq:boundb} equals~\eqref{eq:b} and where~\eqref{eq:boundbplus} equals~\eqref{eq:bplus}. The inequalities yield the desired result. \hfill\IEEEQED

\bibliographystyle{IEEEtran}
\bibliography{IEEEabrv,%
../../References/polarcodes.bib,%
../../References/textbooks.bib}

\end{document}

%% file: abbrevations_processing.tex
\newcommand{\loss}[2][\empty]{\ifthenelse{\equal{#1}{\empty}}{L(#2)}{L_{#1}(#2)}}
\newcommand{\lossrate}[2][\empty]{\ifthenelse{\equal{#1}{\empty}}{L(\mathbf{#2})}{L_{\mathbf{#1}}(\mathbf{#2})}}

\newcommand{\relLoss}[2][\empty]{\ifthenelse{\equal{#1}{\empty}}{l(#2)}{l_{#1}(#2)}}

\newcommand{\dom}[1]{\mathcal{#1}}

\newcommand{\Prob}[1]{\mathrm{Pr}(#1)}

\newcommand{\limn}{\lim_{n\to\infty}}

\newcommand{\infodim}[1]{d(#1)}
